\documentclass[aps,prb,groupedaddress,showpacs,preprint,amssym]{revtex4}
\usepackage{latexsym}
\usepackage{graphicx}
\usepackage{rotating}
\usepackage{longtable} 
\usepackage[usenames]{color}

\begin{document} 

\newcommand{\lsmo}{\ensuremath{\mathrm{La_{0.5}Sr_{1.5}MnO_4}}}
\newcommand{\oo}{($\frac{1}{4}$,$\frac{1}{4}$,0)}
\newcommand{\mo}{($\frac{1}{4}$,$-\frac{1}{4}$,$\frac{1}{2}$)~}
\newcommand{\lsmobi}{\ensuremath{\mathrm{LaSr_{2}Mn_{2}O_7}}~}

\title{Mn L$_{2,3}$ edge resonant x-ray scattering in manganites: Influence of the magnetic state}

\author{N. Stoji\' c,$^{\ 1}$ N. Binggeli,$^{\ 1,2}$ and M. Altarelli$^{\ 1,3}$   }

\affiliation{ $^{\ 1}$  Abdus Salam International Centre for Theoretical Physics,
                         Trieste 34014,
                         Italy}

\affiliation{ $^{\ 2}$ INFM Democritos National Simulation Center,
                       Trieste I-34014, Italy }

\affiliation{ $^{\ 3}$ Sincrotrone Trieste,
                       Area Science Park,
                       34012 Basovizza, Trieste, Italy   }

\date{\today}

\begin{abstract}
We present an analysis of the dependence of the resonant-orbital order and magnetic 
scattering spectra on the spin configuration.
We consider an arbitrary spin direction with respect to the local crystal field axis,  
thus lowering significantly the local symmetry. To evaluate the atomic scattering in this
case, we generalized the Hannon-Trammel formula and implemented it inside the framework
of atomic multiplet calculations in a crystal field. 
For an illustration, we calculate the magnetic and orbital scattering
in the CE phase of \lsmo~ in the cases when the spins are aligned with the crystal lattice vector ${\vec a}$ 
(or equivalently ${\vec b}$)
 and when they are rotated in the $ab$-plane  by 45$^{\circ}$ with respect to this axis. 
Magnetic spectra  differ for the two cases. 
For the orbital scattering,  we show that for the former
configuration there is a non negligible  $\sigma \rightarrow \sigma'$ ($\pi \rightarrow \pi'$)
scattering component, which
vanishes in the 45$^\circ$ case,
while the $\sigma \rightarrow \pi'$ ($\pi \rightarrow \sigma'$) components are similar in the two cases. 
From the consideration of two 90$^\circ$ spin canted 
structures, we conclude there is a significant dependence of the orbital scattering spectra on 
the spin arrangement. 
Recent experiments detected a sudden decrease of the orbital scattering intensity upon increasing
the temperature above the N\' eel temperature in \lsmo. We discuss this behavior considering the effect
of different types of misorientations of the spins on the orbital scattering spectrum.

\end{abstract}

\pacs{61.10.-i,75.10.-b,75.25.+z,71.27.+a}

\maketitle

\section{ Introduction}

The physics of perovskite-type manganites is one of the very exciting areas of the present-day 
solid-state research, displaying abundance of interesting phenomena, of which the most known
is the huge resistivity change upon the application of  magnetic fields, the so-called colossal
magnetoresistance effect\cite{MorAsaKuw96}.
Manganites have rich phase diagrams with several types of magnetic, charge and orbitally 
ordered phases, often exhibiting mesoscale phase coexistence. The physics of these systems is controlled by
a complex interplay of magnetic, orbital, charge and structural degrees of freedom.
Orbital order is characterized by the spatial arrangement of the outermost valence electrons.
Although it was predicted and described theoretically 50 years
ago\cite{Goo55}, it has been recognized as a hidden degree of freedom, because 
of difficulties related to its direct observation. 

The resonant x-ray scattering (RXS)
is a promising tool for the study of orbital order, and, at the same time, convenient for the
study of magnetic and charge ordering. In particular, it can be applied to investigate the correlations
between the orbital order parameter and  the magnetic and charge order parameters.
 The first RXS experiments with the goal of detecting orbital ordering were done 
at the K-edge. At the manganese K-edge, one $1s$ electron is excited into the empty $4p$ band
after absorption of a photon of the resonant energy. This intermediate state
has a finite lifetime, after which the ion emits a photon and relaxes back
to the initial state.
Today it is believed that the observed signal is not sensitive to the onsite orbital ordering of the
$3d$ orbitals, but to band-structure effects including the effect of the Jahn-Teller crystal field
on the 4$p$ states together with the hybridization of the 4$p$ orbitals
with the neighboring oxygen orbitals (also controlled  by the Jahn-Teller distortion)
and with empty 3$d$ orbitals of the neighboring Mn atoms\cite{ElfAniSaw99,BenJolNat99,BenBriPav01}.
Instead, the L$_{2,3}$-edge scattering 
measurements\cite{WilHatRop03,WilSpeHat03,DheMirNad04,ThoHilGre04,WilStoBea05,StaScaMul05}, 
stimulated by the theoretical work of Castleton and Altarelli\cite{CasAlt00},
are sensitive to both the onsite orbital ordering and the Jahn-Teller distortion.
At the Mn L$_{2,3}$ edges, the virtual dipole transition consists of an excitation 
of one $2p$ electron into the $3d$ shell, thus directly probing the orbital 
ordering. 
The specific contributions of the orbital ordering 
and Jahn-Teller distortions must then be disentangled in order to interpret the results of  such experiments. 
This is usually done by comparing the experimental spectra to theoretical simulated spectra.
Previous theoretical work \cite{WilSpeHat03,WilStoBea05} has indicated that the Jahn-Teller distortion affects
almost exclusively the L$_3$ edge. However, to
monitor the effect of a change in the magnetic order on the orbital ordering in particular (or on the
Jahn-Teller distortion), one needs also to know the direct influence of the magnetic configuration on the orbital spectra.
So far, to our knowledge, this effect on the L edge orbital scattering has not been addressed.
Also, obviously, if the spin configuration has a significant influence on the orbital scattering, it
must be properly included in the analysis used to disentangle the effects of the orbital ordering and
Jahn-Teller distortion. 

Recently, Wilkins {\it et al.}\cite{WilStoBea05} reported  measurements of 
the temperature dependence of the orbital scattering
at the Mn L$_{2,3}$ edge in \lsmo~ which displayed a substantial  decrease in the orbital scattering intensity, 
 as the temperature is raised, at
the transition from the antiferromagnetic to paramagnetic phase. Analogous behavior when crossing the N\' eel 
temperature was previously reported in LaMnO$_3$\cite{MurHilGib98}, at the Mn K edge, and in KCuF$_3$\cite{PaoCacSol02,CacPaoSol02} at the Cu K edge (but the
origin may not be the same). More recently, Staub {\it et al.}\cite{StaScaMul05} 
also presented measurements of the temperature dependence
of the orbital scattering at the Mn L$_{2,3}$ edge in \lsmo,  which showed a smaller change at the 
N\' eel temperature. Furthermore, based on a polarization analysis, they showed that the orbital
signal below the N\' eel temperature includes a non negligible contribution of magnetic scattering
from twinned minority ferromagnetic domains.
However, it is still unclear whether this fully explains the temperature dependence of the observed
orbital signal\cite{WilStoBea05,StaScaMul05}. 
In principle, the change in the magnetic configuration at the N\' eel temperature  in the majority
domain
could also affect the spectrum.

In this work, we address the direct influence of the magnetic configuration on the orbital scattering
spectra  for a fixed orbitally ordered configuration. As a prototype system, we focus on \lsmo. We find that the  spin configuration has indeed an influence
on the orbital scattering, but mostly when the spins are noncollinear. For the magnetic scattering
 we find that the spin configuration has a non-negligible influence on the spectrum, and this
even in the collinear CE structure.
Previous studies generally considered spin oriented along the highest local symmetry
axis of the Mn$^{3+}$ crystallographic site. In order to allow an arbitrary orientation of the spin
with respect to the local crystal field axis,
we generalized the Hannon-Trammell formula\cite{HanTraBlu88} for atomic
scattering, derived for spherical local site symmetry,
by a systematic expansion in invariants of decreasing symmetry down to C$_i$. We use these results
to derive the scattering formulae for different spin configurations, including the polarization
dependence.  Finally, we consider various types of  spin misalignments and misorientations
to understand the effect of spin disorder on the orbital spectrum of 
the paramagnetic phase. We separately discuss the effect
of the loss of spin order between the $ab$-planes at temperatures just above T$_N$ and spin misalignments
inside these planes at higher temperatures.

This work is organized as follows: in Section II we describe the system and the theoretical method, 
while in Section III we present the scattering formula in the C$_i$ symmetry, and its applications
towards  the description of the orbital and magnetic scattering
in four special cases of spin arrangement. 
In Section IV, we present the corresponding orbital and magnetic spectra for \lsmo~ and
address the influence of the spin configuration, while in Section V we discuss 
our results in connection with the issue of the
orbital spectrum of the paramagnetic phase. Section VI gives the conclusions of the work, while
two appendices describe some calculational details.

\section{Description of the system and theoretical approach}

\begin{figure}[ht]
  \begin{center}
  \includegraphics[width=9cm]{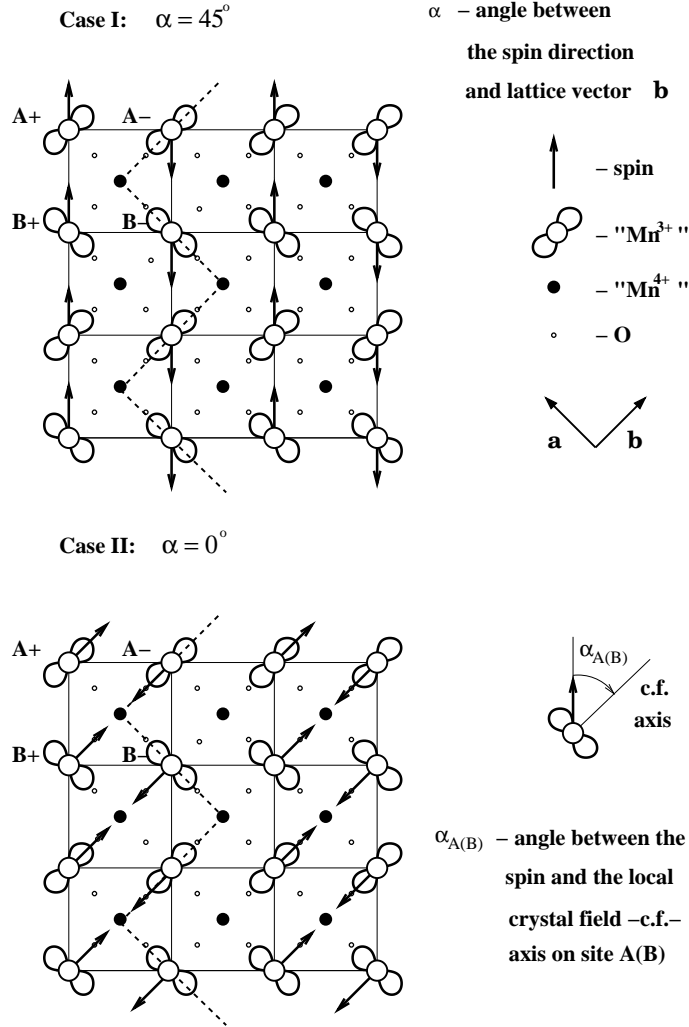}
  \caption{ The MnO$_2$ plane of \lsmo~ representing charge, orbital and spin ordering for two 
different cases of spin orientation in the CE structure. Case I:
the spin and crystal axis ${\vec b}$ enclose an angle $\alpha$ = 45$^{\circ}$ and Case II: $\alpha$ = 0$^{\circ}$.The dashed lines denote the ferromagnetic
spin chains. For  reasons of clarity, the spins on the ``Mn$^{4+}$'' ions are omitted. We assumed the 
$x^2-z^2/y^2-z^2$ type of orbital ordering \cite{WilStoBea05},
in the crystal coordinate system with x, y, and z along the  ${\vec a}$,
 ${\vec b}$, and  ${\vec c}$ axis, respectively. }
  \bigskip
  \label{fig:structure}
  \end{center}
\end{figure} 

\begin{figure}[ht]
  \begin{center}
  \includegraphics[width=10.5cm]{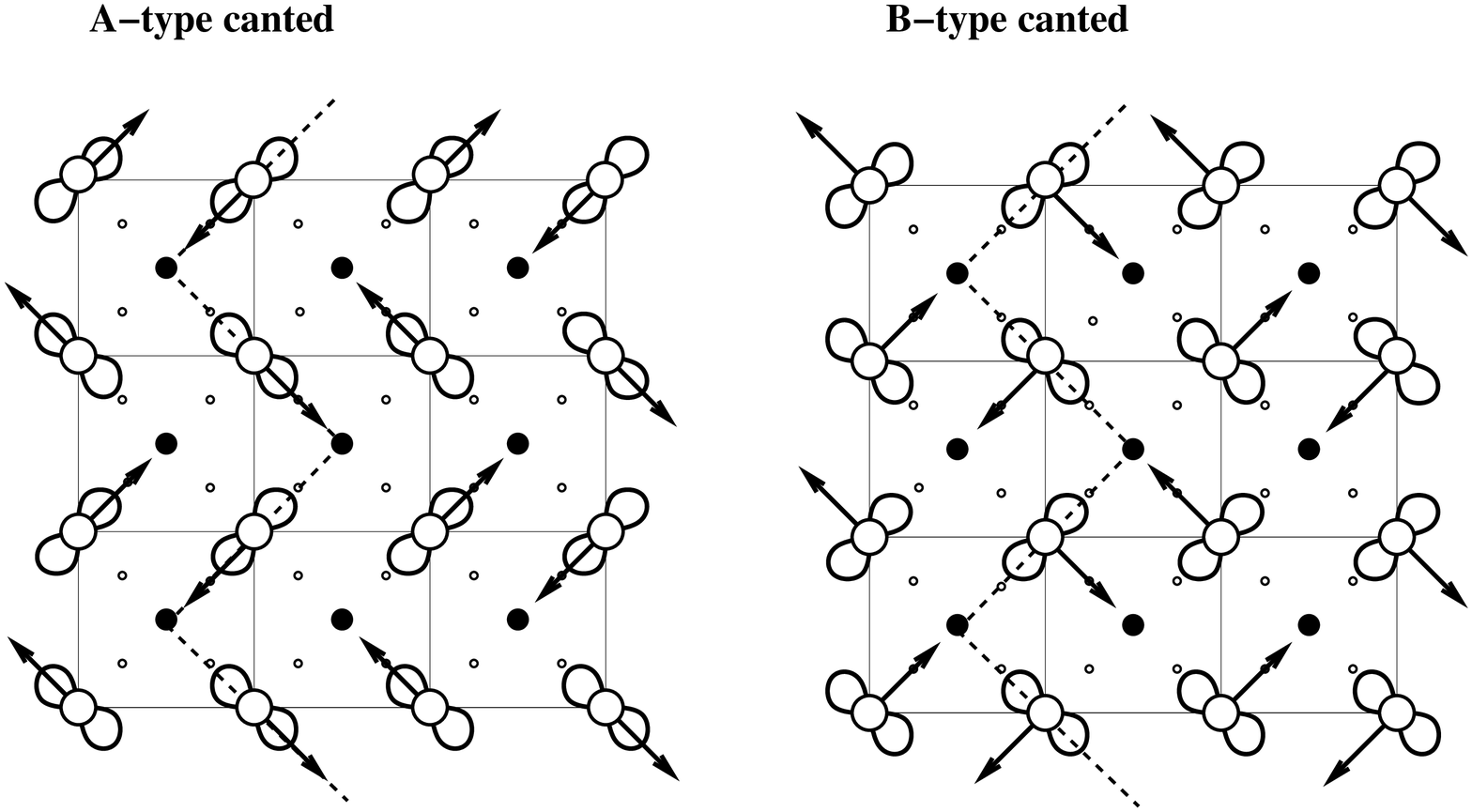}
  \caption{ The MnO$_2$ plane of \lsmo~ representing charge, orbital and spin ordering for two 
hypothetical 90$^\circ$-spin-canted structures.} 
  \bigskip
  \label{fig:canted}
  \end{center}
\end{figure}

In the present work, we focus our attention on the  
half doped manganites with a low temperature CE antiferromagnetic 
structure. As an example, we consider the \lsmo~ compound, 
which displays a variety of phases\cite{LarMecLu05,SteHilWil96}.
The charge and orbitally ordered phase sets on upon cooling at 230~K. 
The  disproportionation of the Mn ions into the formal valences 
Mn$^{3+}$ and Mn$^{4+}$ is usually assumed, while recent experiments\cite{GreHilGib04,HerGarSub04} 
indicate that 
the charge difference in manganite materials is 
much smaller than 1. In the present atomic-based approach,
we  approximate the charge on the manganese ions
by the integral values of 3+ and 4+. 
The four (three) $3d$ electrons of the Mn$^{3+}$ (the Mn$^{4+}$)
 are under the strong influence of the crystal field. 
In the presence of the cubic ligand field, the degenerate $d$ orbitals  
split into triple degenerate $t_{2g}$ and doubly degenerate
$e_g$ levels. Cooperative Jahn-Teller distortions of the MnO$_6$ octahedra further reduce 
the symmetry of the crystal field to $D_{4h}$ at the Mn$^{3+}$ site and thus lift the orbital  degeneracy.
$t_{2g}$ level splits into $d_{xy}$ and doubly-degenerate $d_{xz, yz}$ orbitals, and $e_g$ into
orbitals of the $d_{3z^2-r^2}$ and $d_{x^2-y^2}$ type.
Only one of the $e_g$ orbitals is occupied on the  Mn$^{3+}$
ions, leading to orbital ordering, {\it i.e.} an alternate arrangement of orbitals
oriented along the ${\vec a}$ and ${\vec b}$ axes on neighboring  Mn$^{3+}$
A and B sites (see Fig.~\ref{fig:structure}).
The antiferromagnetic phase, characterized as the CE type structure, exists under 115~K.
In this phase, the spins are lying in the $ab$-plane\cite{note2},
displaying ferromagnetic ordering along spin-chains in the $ab$-plane (see Fig.~\ref{fig:structure}:
${\rm A^--B^--A^-}$ and ${\rm A^+-B^+-A^+}$ spin chains, ``+'' and ``-'' refer to the spin direction)
and antiferromagnetic ordering between the spin chains both 
in the $ab$-plane and along the $c$-axis.

Figure~\ref{fig:structure} presents two possible cases of spin orientation in the CE structure, which 
we will consider in the present work. In Case I, the spins are parallel to the ${\vec a + \vec b}$ axis,
corresponding to an angle $\alpha=45^\circ$ between the spin direction and the crystal ${\vec b}$ axis.
In Case II, instead, the spins are parallel to the ${\vec b}$ axis ($\alpha = 0^\circ$). 

Having in mind the paramagnetic phase and the effect of spin misorientations on the orbital
spectrum, we will also consider two 90$^\circ$-spin-canted structures, which are illustrated in 
Fig.~\ref{fig:canted}. These structures, ``A-type canted'' and ``B-type canted'', are built
using local atomic configurations of Case II in Fig~\ref{fig:structure}. The A-type canted is
obtained using the Mn$^{3+}$ A$^{(+,-)}$ configurations (spin parallel to the orbital) and the
B-type canted using the  Mn$^{3+}$ B$^{(+,-)}$ configurations (spin perpendicular to the orbital),
with possible 90$^\circ$ rotations of these atomic building blocks. We note that, for the orbital
scattering, only Mn$^{3+}$ sites need to be considered. Also, for the magnetic scattering of the
CE structures, we will consider (as in the experiments\cite{WilSpeHat03,WilStoBea05}) only superlattice
reflections which selectively probe the Mn$^{3+}$ spins contribution.

We based our calculations of the RXS spectra  on  atomic multiplet 
formulation in a crystal field\cite{Gro04,LaaTho91}. 
Cowan's atomic multiplet 
program \cite{Cow68,Cow81} provides the Hartree-Fock 
values of the radial Coulomb (Slater) integrals  $F^{0,2,4}(d,d)$, $F^2(p,d)$,
$G^{1,3}(p,d)$ (direct and exchange contributions) and the
spin-orbit interactions $\zeta(2p)$ and $\zeta(3d)$ for an isolated Mn$^{3+}$  ion
at  0~K. They are given in Table~\ref{HartreeFockvalues}.  It is known that the strengths of 
the Coulomb integrals
in a solid are reduced due to  the screening effects to about $\sim$70-80\% of the atomic values\cite{GroFugTho90}. 
The largest of these integrals are  nearly 
an order of magnitude greater than the crystal field splittings.  This fact
justifies our choice of the atomic multiplet approach in a crystal field.
\renewcommand{\baselinestretch}{1}
\begin{table}[ht]
\bigskip
\begin{center}
\begin{tabular}{|l c c c c c c c| }
\hline
\hline
$ $ & $F^{2}(d,d)$  &  $F^{4}(d,d)$ &   $F^{2}(p,d)$ & $G^{1}(p,d)$ &$G^{3}(p,d)$  &$\zeta(2p)$ &$\zeta(3d)$
    \\
\hline
$3d^4$  & $11.415$ & $7.148$ & $-$ & $-$ & $-$ & $-$ & $0.046$\\
$2p^53d^{5}$ &  $12.210$ & $7.649$  & $6.988$ & $5.179$ & $2.945$ & $7.467$ & $0.059$ \\
\hline
\hline
\end{tabular}
\caption{The Hartree-Fock values for the ground and excited states of $\rm Mn^{3+}$ given in eV.
 The $p$-shell spin-orbit parameter $\zeta(2p)$ has been increased
by 9~\% from the Hartree-Fock value to correspond to the experimental value \cite{ThoAttGul01}. }
\label{HartreeFockvalues}
\end{center}
\end{table}

The treatment of the crystal field in this model is based on symmetry considerations.
Starting from the transition matrix elements evaluated in the spherical point group, 
we lower the symmetry by the inclusion of a crystal field, following a ``branching'' of point
groups.
For example, the following ``branching'' can be generated:
 $O_3 \supset O_h \supset D_{4h} \supset D_{2h} \supset C_i$ \cite{But81}.
For the calculation of the
matrix elements in a given point group, the Wigner-Eckart
theorem is used. In practice, this theorem is applied through
a multiplication of the matrix element evaluated 
in the O$_3$ point
group by a chain of 2-j and 3-j coefficients for each branching. 
This is implemented
in a crystal field code, which gives the energy spectrum and the
dipole transition matrix elements necessary for the calculation of the scattering spectra.
The values of the cubic  (${\rm X^{400}}$) and
tetragonal (${\rm X^{220}}$) crystal field parameters are obtained from a fit to 
the experimental RXS spectra. In the present case, we will use the values obtained
in Ref.~\onlinecite{WilStoBea05} for \lsmo. These values are ${\rm X^{400}}=5.6$~eV and
${\rm X^{220}}=3.75$ eV, and correspond to an $x^2-z^2/y^2-z^2$ type of orbital ordering (with
the x, y, and z axes along ${\vec a},{\vec b}$ and ${\vec c}$, respectively).
All the Slater integrals were scaled down to 75~\% of their atomic values\cite{GroFugTho90,AbbGroFug92}.
Furthermore, in order to simulate the effects of the spin ordering on an isolated ion
(superexchange and direct exchange interactions),
we introduce a magnetic field acting on the spin of the atom. This field splits
additionally the $S=2$ quintet into $S_z=-2,-1, \dots,2$ levels. Inclusion of the magnetic 
field  favors the $S_z=-2$ level as the ground state of the atom. 
The value of the Zeeman energy is set to 0.02~eV.

\section{Scattering formula}

In the first part of this section we describe atomic scattering 
for a general orientation of the spin axes with respect to the local crystal field axes.
Thereafter, we apply the atomic scattering formula to find expressions for the orbital
scattering of the four different spin structures considered
and the magnetic scattering for the two different orientations of the spins in the antiferromagnetic
 CE-phase.

\subsection{Systematic expansion of the atomic scattering amplitude in invariants of decreasing symmetry}

To calculate a diffraction spectrum,  we first evaluate the atomic scattering from a single site and then add up the
contributions of the atoms in the whole crystal. The type of order is selected by the scattering vector
$\vec{q}$ which describes the periodicity of the orbital, magnetic or charge order.
Depending on the relative orientation of the local 
crystal field axis and the spin direction, the local symmetry can change.
In the case in which the crystal and spin axes are collinear,
the site symmetry  corresponds to the one of the crystal field. 
We use a tetragonal crystal field\cite{note4}, described by the $D_{4h}$ point group. 
If the spin direction, however, differs from the crystal field axis, the local symmetry
can be as low as that of the $C_i$ point group.

Following the approach by Carra and Thole\cite{CarTho94}, we  write the scattering amplitude 
as a linear combination of the product of pairs of tensors of increasing rank, which transform according to the 
irreducible representations of the spherical group SO$_3$. To lower the symmetry to the  $C_i$ point group,
we need to branch each irreducible representation into subgroup of irreducible representations, following
the chain: $O_3 \supset O_h \supset D_{4h} \supset C_{4h} \supset C_{2h} \supset C_i$. Only the 
totally symmetric representations  contribute to the atomic scattering amplitude. If we stop the branching at the
C$_{4h}$ point group, we find for the atomic scattering amplitude the following expression:
\begin{equation}
\begin{array}{lll}
f^{\rm E_1}_{\rm res,C_{4h}}&=&\frac{3}{4} \lambda \Big\{ {(\bf \hat e}_f^* \cdot {\bf \hat e}_i)(F_{1;1}+F_{-1;-1}) - 
 i \left[({\bf \hat e}_f^* \times {\bf \hat e}_i)\cdot{\bf \hat  z}\right] (F_{1;1}-F_{-1;-1})+  \\ 
& & ({\bf \hat e}_f^* \cdot{\bf \hat  z})({\bf \hat e}_i \cdot {\bf \hat z}) (2 F_{0;0} - F_{1;1}-F_{-1;-1})  \Big\},
\end{array}
\label{eq:scattering_C4h}
\end{equation}
where $\lambda$ represents a scattering coefficient\cite{CarTho94}, $ {\bf \hat e}_i( {\bf \hat e}_f)$ is 
the incoming (outcoming) photon polarization, and $ F_{\rm m;m'}$ is  defined as:
\begin{equation}
F_{{\rm m;m'}}=\sum_n \frac{\langle 0|J^{1\dag}_{\rm m}|n\rangle \langle n|J^{1}_{\rm m'}|0\rangle}{E_0-E_n+\hbar \omega + i\Gamma /2},
\label{eq:defineF}
\end{equation}
where m and m$'$ denote polarization states and $J^1_m$ are the electric dipole operators defined in spherical coordinates. $|0\rangle$ represents the ground state with energy $E_0$ and
$|n\rangle$ an intermediate state with energy $E_n$. The photon energy is $\hbar \omega$ and
$\Gamma$ represents the broadening due to the core-hole lifetime.
The local coordinate system is chosen in such a 
way that the z-axis corresponds to the highest local symmetry axis, which is also the
quantization (spin) direction.
  Eq. \ref{eq:scattering_C4h} is known as the Hannon-Trammel formula\cite{HanTraBlu88}, which was derived for
the spherical symmetry, and is applicable up to local symmetries as low as that
of the  $C_{4h}$ point group.
 Lowering the symmetry to the C$_{2h}$ point group,
the following expression for the atomic scattering amplitude is found:
\begin{equation}
\begin{array}{lll}
f^{\rm E_1}_{\rm res,C_{2h}}&=& f^{\rm E_1}_{\rm res,C_{4h}} + 
\frac{3}{4} \lambda \Big\{  \left[ ({\bf \hat e}_f^* \cdot {\bf \hat y})({\bf \hat e}_i \cdot{\bf \hat y})- ({\bf \hat e}_f^* \cdot {\bf \hat x})({\bf \hat e}_i \cdot{\bf \hat  x})\right] (F_{-1;1}+F_{1;-1}) + \\ 
& & i\left[ ({\bf \hat e}_f^* \cdot {\bf \hat x})({\bf \hat e}_i \cdot{\bf \hat y})+ ({\bf \hat e}_f^* \cdot{\bf \hat y})({\bf \hat e}_i \cdot {\bf \hat x})\right] (F_{-1;1}-F_{1;-1}) \Big\},
\end{array}
\end{equation}
where {${\bf \hat x}$, ${\bf \hat y}$, and ${\bf \hat z}$ are unit vectors corresponding to a local 
right Cartesian coordinate system, 
with the
z-axis defined as above. 
Finally, assuming no symmetry operation on the site, except the inversion, we obtain four additional
terms in the description of the atomic scattering amplitude in the C$_i$ point group:
\begin{equation}
\begin{array}{lll}
f^{\rm E_1}_{\rm res,C_{i}}= f^{\rm E_1}_{\rm res,C_{2h}} + \frac{3}{4} \lambda \Big\{ 
&-i&\frac{\sqrt{2}}{2}  \left[ ({\bf \hat e}_f^* \cdot{\bf \hat y})({\bf \hat e}_i \cdot {\bf \hat z})- ({\bf \hat e}_f^* \cdot{\bf \hat z})({\bf \hat e}_i \cdot {\bf \hat y})\right] (F_{1;0}+ F_{-1;0}+F_{0;1}+F_{0;-1}) + \\ 
& & \frac{\sqrt{2}}{2} \left[ ({\bf \hat e}_f^* \cdot{\bf \hat x})({\bf \hat e}_i \cdot {\bf \hat z})-({\bf \hat e}_f^* \cdot{\bf \hat z})({\bf \hat e}_i \cdot {\bf \hat x})\right] (-F_{1;0}-F_{0;-1}+F_{-1;0} + F_{0;1}) + \\ 
&-i& \frac{\sqrt{2}}{2} \left[ ({\bf \hat e}_f^* \cdot{\bf \hat y})({\bf \hat e}_i \cdot{\bf \hat z})+ ({\bf \hat e}_f^* \cdot{\bf \hat z})({\bf \hat e}_i \cdot{\bf \hat y})\right] (F_{1;0}+ F_{-1;0}-F_{0;1}-F_{0;-1}) + \\ 
& & \frac{\sqrt{2}}{2} \left[ ({\bf \hat e}_f^* \cdot{\bf \hat x})({\bf \hat e}_i \cdot{\bf \hat z})+ ({\bf \hat e}_f^* \cdot{\bf \hat z})({\bf \hat e}_i \cdot{\bf \hat x})\right] (-F_{1;0}+F_{0;-1}+F_{-1;0} - F_{0;1})\Big\},
\label{eq:Ci-scattering}
\end{array}
\end{equation}
where ${\bf \hat x}$, ${\bf \hat y}$, and ${\bf \hat z}$ are unit vectors of any right Cartesian 
coordinate system, but it is convenient to choose, also in this
case, the local z-axis along the spin direction.

\subsection{Orbital order scattering}

\begin{figure}[h]
  \begin{center}
  \includegraphics[width=7.7cm]{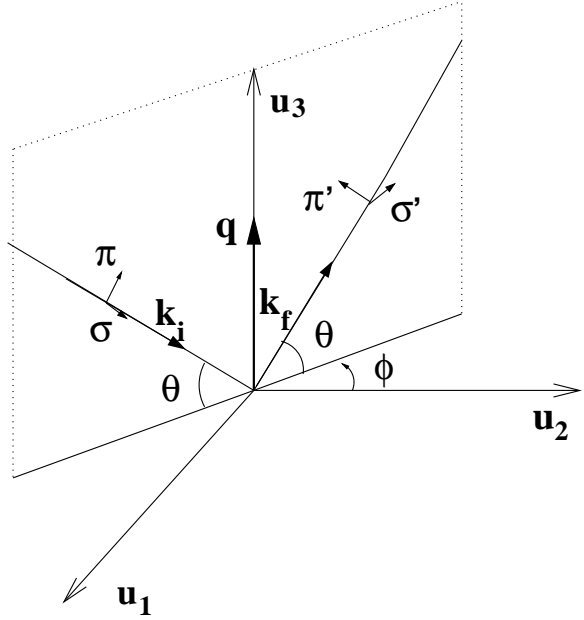}
  \caption{ The scattering plane with the definition of the scattering and azimuthal angles $\theta$ and $\phi$. 
${\vec k}_i ({\vec k}_f)$ is the incident (outcoming) photon wavevector, ${\vec q}={\vec k}_f-{\vec k}_i$. 
For the case of orbital order scattering with ${\rm \vec{q}^{OO}=}$\oo, the reference axes are:
 ${\rm \vec{u}_1=\frac{1}{\sqrt{2}}[\bar 1 1 0]}$, 
${\rm \vec{u}_2=[0 0 1]}$ and ${\rm \vec{u}_3=\frac{1}{\sqrt{2}}[1 1 0]}$ in the crystal coordinate system ${\vec a}$, ${\vec b}$, ${\vec c}$. 
For the magnetic scattering, ${\rm \vec{q}^{AF}=}$\mo 
and the following reference axes are used: ${\rm \vec{u}_1=\frac{1}{\sqrt{3}}[ 1 \bar 1 \bar 1]}$, ${\rm \vec{u}_2=\frac{1}{\sqrt{2}}[1 1 0]}$ and
${\rm \vec{u}_3=\frac{1}{\sqrt{6}}[1 \bar 1 2]}$. }
  \bigskip
  \label{fig:scatter}
  \end{center}
\end{figure} 

In tetragonal CE structures, the wave vectors probing the antiferro orbital ordering 
are: ${\vec q}^{{\rm OO}}=(\frac{1+2n}{4}-\frac{m}{2},\frac{1+2n}{4}+\frac{m}{2},l)$ 
in units of $(\frac{2\pi}{a},\frac{2\pi}{a},\frac{2\pi}{c})$, where $n$, $m$, and $l$ are integers 
and $a,c$ are the lattice constants.
Taking into account the structure factors, the scattering amplitude for orbital wave vectors is proportional
to the following combination of the atomic scattering amplitudes at the A$^{(+,-)}$ and B$^{(+,-)}$ sites: 
$f^{A+} + f^{A-} - f^{B+} - f^{B-}$. In the soft x-ray experiments\cite{WilSpeHat03,DheMirNad04,StaScaMul05} the orbital order wave vector was taken
as ${\vec q}^{{\rm OO}}=$\oo, and this is the wave vector we will consider for the polarization analysis presented
in the next parts of this section. The equations will be presented using as reference  system for the
components of the atomic scattering tensors $F_{m,m'}$ the local coordinate system
in which the z-axis remains in the spin direction, x lies in the $ab$-plane and the y-axis is perpendicular to the $ab$-plane.

\subsubsection{Case I: $\alpha=45^\circ$}

For the case when $\alpha=45^\circ$, it is useful to note that starting from the configuration at site A+, 
in Fig.~\ref{fig:structure},
and using symmetry operations (rotation by 180$^\circ$ about the ${\vec c}$ or (${\vec a}+{\vec b})$ axis)
one can obtain the configurations at the remaining three sites. 
This simplifies the numerical part of the calculation,
because it suffices to perform an atomic calculation for just one site. 
The local symmetry of each site is described by the C$_i$ point group. 
From inspection of Eq. \ref{eq:Ci-scattering}
it can be seen that the only nonvanishing terms of the orbital order scattering are the ones including 
$({\bf \hat e}_f^* \cdot {\bf \hat x})({\bf \hat e}_i \cdot {\bf \hat z})$
and $({\bf \hat e}_f^* \cdot {\bf \hat z})({\bf \hat e}_i \cdot{\bf \hat  x})$ polarization factors. 
Furthermore,  the second term in the curly brackets in
 Eq. \ref{eq:Ci-scattering} vanishes, in our case, as the matrix elements 
in Eq. \ref{eq:defineF} are real.\cite{note1}
After evaluation of the polarization
factor we obtain the following expression for the orbital order scattering amplitude:
\begin{equation}
f^{\rm OO}_{\rm res}=-\sqrt{2} \cos\theta \cos\phi \left( -F_{1;0}^{A+}+F_{0;-1}^{A+}+F_{-1;0}^{A+}-F_{0;1}^{A+}\right),
\label{eq:OO_45}
\end{equation}
in the case of $\sigma  \rightarrow \pi'$ or $\pi \rightarrow \sigma'$ polarization. In Eq.~\ref{eq:OO_45}
the scattering and azimuthal angles $\theta$ and $\phi$ are defined as shown in Fig. \ref{fig:scatter}, using as
reference axes  ${\rm \vec{u}_1=\frac{1}{\sqrt{2}}[\bar 1 1 0]}$, ${\rm \vec{u}_2=[0 0 1]}$ 
and ${\rm \vec{u}_3=\frac{1}{\sqrt{2}}[1 1 0]}$ in the crystal coordinate system ${\vec a}$, ${\vec b}$, ${\vec c}$. The scattering amplitude
vanishes in  the case of  $\sigma  \rightarrow \sigma'$ or  $\pi \rightarrow \pi'$ polarization.
The F term in parenthesis of Eq.~\ref{eq:OO_45} can be written in Cartesian coordinates as $2F_{xz}$.

\subsubsection{Case II: $\alpha=0^\circ$}

In the case when the angle $\alpha$ is different from 45$^\circ$, the sites A and B are inequivalent, since
the spin encloses two different angles, $\alpha_A$ and $\alpha_B$, 
with the local crystal field axis at sites A and B. 
Case II, with $\alpha=0^\circ$ in Fig.~\ref{fig:structure}, 
illustrates the situation with the highest local site symmetry, when
the spin is collinear with the crystal field
axis at site B ($\alpha_B$ = 0$^\circ$), and perpendicular to the local crystal field
axis at site A ($\alpha_A = 90^\circ$).
 The local site symmetries for the B and A sites are C$_{4h}$ and C$_{2h}$ respectively.
For the evaluation of orbital scattering, we add the contributions from the four sites in the
same manner as we did in the previous case. But, this time, we 
need to do atomic calculations for two different sites, as there is no symmetry operation relating sites A and B.
For the same reason, there will be no cancellation of contributions from various terms, which we had 
previously, and the expression describing orbital order scattering amplitude will have the following form for 
$\sigma$ polarized incident light:
\begin{equation}
\begin{array}{lll}
f^{{\rm OO, \sigma_{inc}}}_{\rm res}&= &\left\{ \begin{array}{ll} \frac{1}{2}\cos^2\phi & \\  \frac{1}{2}\cos\phi (\sin\theta \sin\phi -\cos\theta) & \end{array}\right\}   \left[2 (F_{0;0}^{A+} - F_{0;0}^{B+} )\right] + \\   
& & \left\{ \begin{array}{ll}\sin^2\phi + \frac{1}{2}\cos^2\phi  & \\ \frac{1}{2}\cos\theta \cos\phi -\frac{1}{2}\cos\phi \sin\theta \sin \phi & \end{array}\right\} \left[F_{-1;-1}^{A+}+F_{1;1}^{A+}-(F_{-1;-1}^{B+}+F_{1;1}^{B+})\right] + \\
& & \left\{ \begin{array}{ll} \frac{1}{2}\cos^2\phi -\sin^2\phi & \\ \frac{1}{2}\cos\phi(3\sin\theta \sin\phi + \cos\theta ) & \end{array}\right\}  \left[-F_{-1;1}^{A+}-F_{1;-1}^{A+}+(F_{-1;1}^{B+}+F_{1;-1}^{B+})\right],
\label{eq:OOscat0-90}
\end{array}
\end{equation}
where the upper expression in the curly brackets represents the $\sigma  \rightarrow \sigma'$ polarization part and
the bottom one the $\sigma  \rightarrow \pi'$ contribution. The superscripts A and B denote 
the two different sites. The second term on the right hand side of Eq.~\ref{eq:OOscat0-90} 
is proportional to $F_{xx}+F_{yy}$ in the local Cartesian coordinates
and the third one to $F_{xx}-F_{yy}$.
The angles $\theta$ and $\phi$ are defined in Fig \ref{fig:scatter}. 
To simplify the polarization factors, we evaluate the polarization dependence for the azimuthal angle, using 
the value  $\phi=$ 0$^\circ$, as in the experiment (Ref.~\onlinecite{WilSpeHat03}). 
Taking this into account, Eq. \ref{eq:OOscat0-90} can
be rewritten in a simpler form:
\begin{equation}
\begin{array}{lll}
f^{{\rm OO, \sigma inc}}_{\rm res} &=&  \left\{ \begin{array}{ll} 1 \\  0&\end{array}\right\} \left[F_{xx}^{A+} + F_{zz}^{A+}-(F_{xx}^{B+} + F_{zz}^{B+})\right] +   \left\{ \begin{array}{ll} 0 \\  \cos\theta &\end{array}\right\}  \left[F_{xx}^{A+} - F_{zz}^{A+}-(F_{xx}^{B+} - F_{zz}^{B+})\right],
\label{eq:OOscat0-90_cart}
\end{array}
\end{equation}
with $F_{xx}^{A(B)+}$ and $F_{zz}^{A(B)+}$ in the local Cartesian coordinates.
We note the important fact that in the $\alpha=0^\circ$ case, 
the $\sigma  \rightarrow \sigma'$ polarization factor is nonvanishing,
in contrast to the $\alpha=45^\circ$   case. 

In  the case of $\pi$ incident polarization, the scattering amplitude reads:
\begin{equation}
\begin{array}{lll}
f^{{\rm OO, \pi_{inc}}}_{\rm res}&= &\left\{ \begin{array}{ll} \frac{1}{2}(\cos^2\theta - \sin^2\theta\sin^2\phi)  & \\  -\frac{1}{2}\cos\phi (\sin\theta \sin\phi +\cos\theta) & \end{array}\right\}   \left[2 (F_{0;0}^{A+} - F_{0;0}^{B+} )\right] + \\   
& & \left\{ \begin{array}{ll}\frac{1}{2}\cos^2\theta - \sin^2\theta(\cos^2\phi+\frac{1}{2}\sin^2\phi)  & \\ \frac{1}{2}\cos\phi(\cos\theta  + \sin\theta\sin\phi) & \end{array}\right\} \left[F_{-1;-1}^{A+}+F_{1;1}^{A+}-(F_{-1;-1}^{B+}+F_{1;1}^{B+})\right] + \\
& & \left\{ \begin{array}{ll}\frac{1}{2}\cos^2\theta + \sin^2\theta(\cos^2\phi-\frac{1}{2}\sin^2\phi)  & \\ \frac{1}{2}\cos\phi(-3\sin\theta \sin\phi + \cos\theta ) & \end{array}\right\}  \left[-F_{-1;1}^{A+}-F_{1;-1}^{A+}+(F_{-1;1}^{B+}+F_{1;-1}^{B+})\right],
\label{eq:OOscat0-90pi}
\end{array}
\end{equation}
where the upper (lower) expression in the curly brackets corresponds to the $\pi \rightarrow \pi'$ ($\pi \rightarrow \sigma'$)
polarization channel. Considering again $\phi=0^\circ$,
we see that  all three terms on the right hand side of Eq.~\ref{eq:OOscat0-90pi} 
contribute to the $\pi \rightarrow \pi'$ scattering, while 
 the expression for the $\pi \rightarrow \sigma'$ scattering is the same as in the case of the 
 $\sigma$-polarized incident light.

\subsubsection{Canted structures}

The orbital scattering amplitude for the two canted structures in Fig.~\ref{fig:canted} 
can be described by the following expression with the $F$'s in the local  Cartesian coordinates:
\begin{equation}
\begin{array}{lll}
f^{{\rm OO, \sigma inc}}_{\rm res} &=&     \left\{ \begin{array}{ll} 0 \\  \cos\theta &\end{array}\right\}  \left[F_{xx}^{A+} - F_{zz}^{A+}-(F_{xx}^{B+} - F_{zz}^{B+})\right],
\label{eq:cantedOO}
\end{array}
\end{equation}
assuming as above $\phi=0^\circ$, and using the same convention as in Eqs.~\ref{eq:OOscat0-90} and \ref{eq:OOscat0-90_cart} 
for the polarization channels.
For incident $\sigma$ polarization, the scattering takes place 
exclusively in the  $\sigma \rightarrow \pi'$ channel.

\subsection{Magnetic scattering}

We performed a similar analysis for the magnetic scattering of the antiferromagnetic CE structures, considering the
wave vector $\vec{q}^{AF}$=\mo which probes the antiferromagnetic order of the Mn$^{3+}$ atoms and was used in the 
experiment (Ref.~\onlinecite{WilSpeHat03}).
The corresponding magnetic scattering amplitude is proportional to: 
$f_{AF}=f^{A+} + f^{B+} - f^{A-} - f^{B-}$.
In the case $\alpha = 45^\circ$ 
the only contributing term is the second term on the right-hand side of Eq.~\ref{eq:scattering_C4h}, 
and, therefore, the antiferromagnetic scattering amplitude
for incident $\sigma$-polarized light can be calculated from the expression:
\begin{equation}
f^{\rm AF, \sigma inc}_{\rm res}=  2 \cos\theta \cos\phi  \left(F_{1;1}^{A+}-F_{-1;-1}^{A+}\right),
\label{eq:AF_45}
\end{equation}
and in the case $\alpha = 0^\circ$:
\begin{equation}
f^{\rm AF, \sigma inc}_{\rm res}=  \left[\cos\theta ( \frac{\sqrt{2}}{2} \cos\phi +\frac{\sqrt{3}}{3} \sin\phi) + \frac{\sqrt{6}}{6}\sin\theta\right] \left(F_{1;1}^{A+}-F_{-1;-1}^{A+} +F_{1;1}^{B+}-F_{-1;-1}^{B+} \right).
\label{eq:AF_0}
\end{equation}
The angles $\theta$ and $\phi$ have been defined as shown in Fig. \ref{fig:scatter}, with the following
definitions of the reference axes: ${\rm \vec{u}_1=\frac{1}{\sqrt{3}}[ 1 \bar 1 \bar 1]}$, ${\rm \vec{u}_2=\frac{1}{\sqrt{2}}[1 1 0]}$ and
${\rm \vec{u}_3=\frac{1}{\sqrt{6}}[1 \bar 1 2]}$ in the crystal coordinate system 
${\vec a}$, ${\vec b}$, ${\vec c}$.  There is no $\sigma  \rightarrow \sigma'$ contribution and the
polarization factor in Eqs. \ref{eq:AF_45} and \ref{eq:AF_0} is evaluated for the $\sigma \rightarrow \pi'$ polarization.
In the local Cartesian coordinates, the $F-$factor from Eqs.~\ref{eq:AF_45} and \ref{eq:AF_0} can be expressed as $F_{xy}-F_{yx}$.

\section{Influence of the spin orientation on the scattering spectra}

In Fig.~\ref{fig:0oo}, we compare the calculated orbital order spectra of the
CE structures with the two different spin orientations,  $\alpha= 0^{\circ}$ and
$\alpha= 45^{\circ}$, displayed in Fig.~\ref{fig:structure}.
We present the spectra for $\sigma$ and $\pi$ incident polarization
evaluated using the equations derived in the previous section.
In the case of the $\sigma$ incident polarization, 
panel~a) displays the spectrum for the $\sigma  \rightarrow \sigma'$ polarization, panel~b)
 for the   $\sigma  \rightarrow \pi'$ polarization, while the  simulated 
experimental situation without a polarization analyzer for the outcoming photons is presented in  panel~c). 
Similarly, for the $\pi$ incident polarization, we present the $\pi  \rightarrow \pi'$
scattering channel on panel~a),  $\pi  \rightarrow \sigma'$ on panel~b) and 
the average of the two in panel~c).
\begin{figure}[ht]
  \begin{center}
  \includegraphics[width=7.7cm, angle=0]{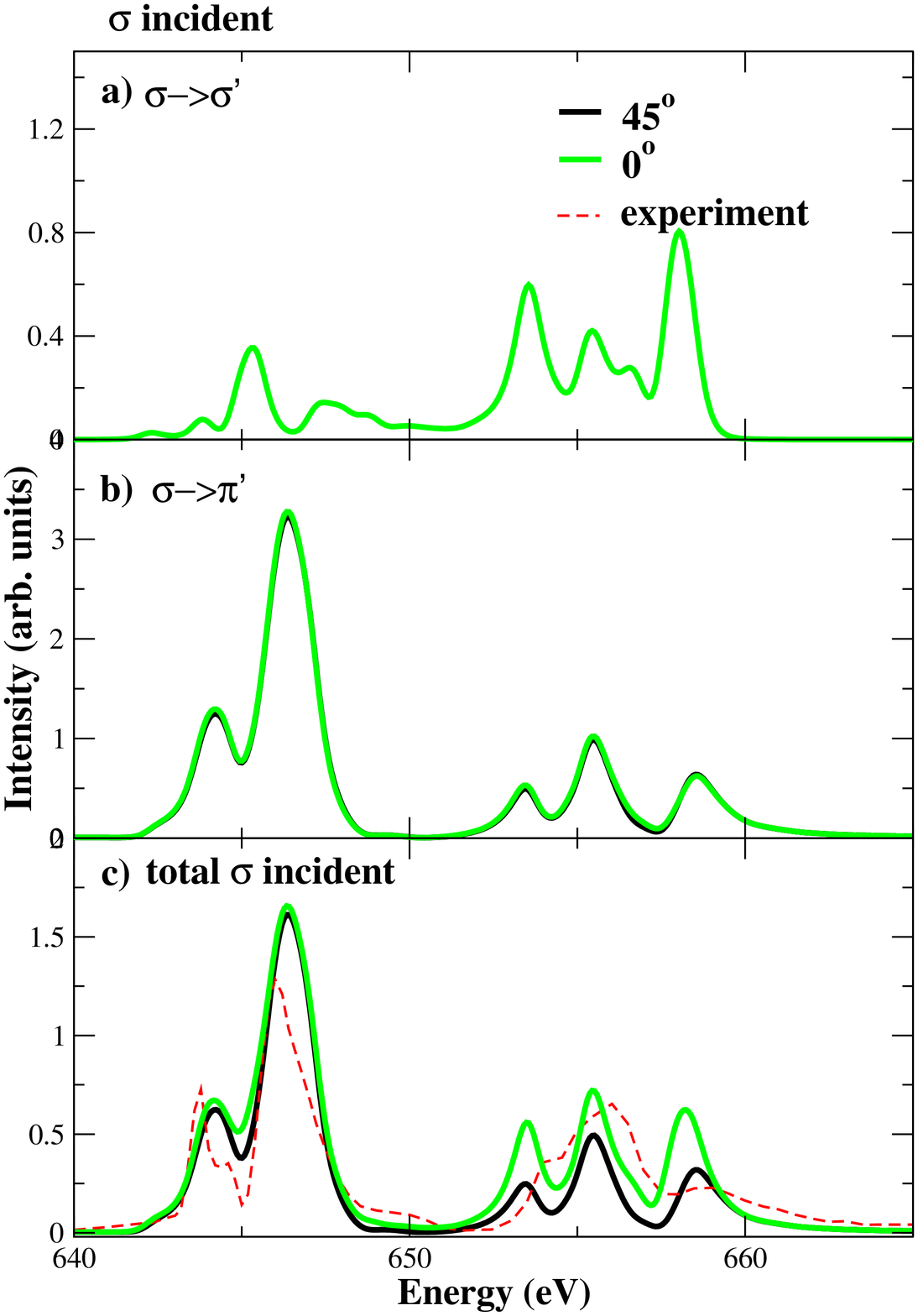}
 \includegraphics[width=7.7cm, angle=0]{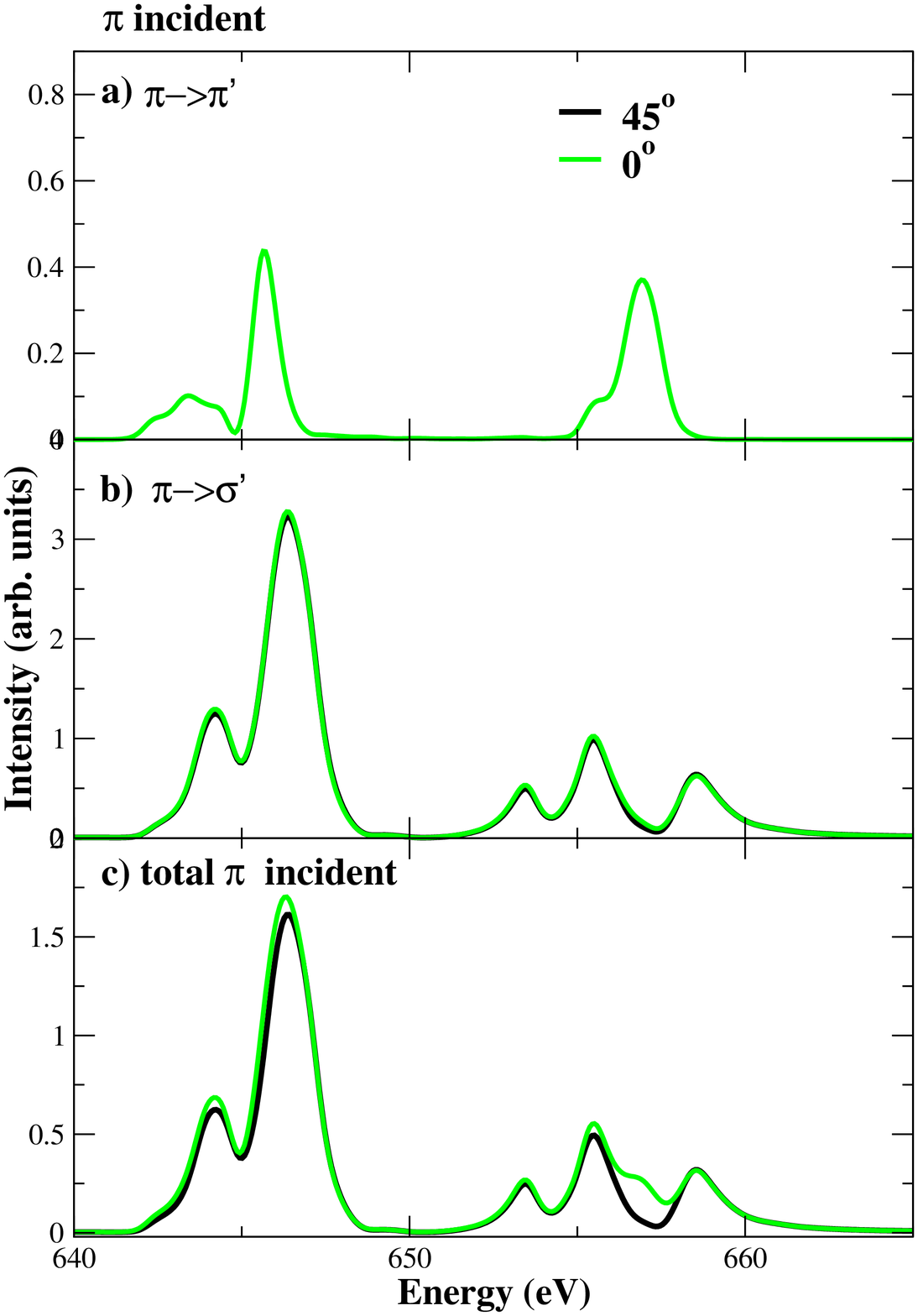}
  \caption{ (Color online) Theoretical orbital order spectra of the CE structures with the two different spin orientations 
$\alpha =  45^\circ$  (black line) and 0$^{\circ}$ (light green line) calculated for $\sigma$ and $\pi$ incident polarizations.
 Different panels represent the polarization
channels: for the $\sigma$ incident a) $\sigma  \rightarrow \sigma'$, b) $\sigma  \rightarrow \pi'$ and c) total, and for the $\pi$ incident a) $\pi \rightarrow \pi'$, b) $\pi \rightarrow \sigma'$ and c) total.
 The experimental spectrum (dashed red line) for $\sigma$ incident polarization
 is reproduced from Ref. \onlinecite{WilStoBea05}.  }
  \bigskip
  \label{fig:0oo}
  \end{center}
\end{figure}
The polarization dependence was calculated for the actual experimental angles:  
$\theta=$ 62.9$^\circ$ and $\phi=$ 0$^\circ$ \cite{WilSpeHat03}.
We used $\Gamma = 0.5$~eV  for the broadening due to the core-hole lifetime and the scattering intensity was 
convoluted with a Gaussian of width 0.1~eV to simulate the experimental (energy) resolution. 

The most significant difference between the spectra for the two angles $\alpha$
are the nonvanishing  $\sigma  \rightarrow \sigma'$ and $\pi \rightarrow \pi'$  contributions
for the 0$^{\circ}$ alignment. For this angle and in the case of the $\sigma$ incident polarization,
 the scattering for the photon energies of the
L$_3$ edge is  very small compared to the scattering at the L$_2$ edge.
For the $\sigma  \rightarrow \pi'$ polarization,
the spectra for 0$^\circ$ and 45$^\circ$ alignments are identical, both in shape and magnitude.
Similarly, for the $\pi$ incident polarization, the $\pi \rightarrow \pi'$  contribution is much
smaller than for the  $\pi  \rightarrow \sigma'$ scattering channel, the spectrum of which is
again identical to the one obtained for $\alpha=45^\circ$.
This can be explained by considering a  linear expansion of $F^{A(B)}$ around the angle $\alpha=45^\circ$. 
It can be shown (the details are given in Appendix B) that the first order corrections of four different
$F$ terms in Eq.~\ref{eq:OOscat0-90_cart} cancel exactly in the case of collinear spins, {\it i.e.}
when the angle $\alpha_A$ at the site A
can be expressed as $\alpha_A=90^\circ - \alpha_B$, leaving just the zeroth order term which equals
the scattering amplitude in the case of 45$^\circ$.
In the situation where both polarization channels contribute with the same weight, 
[panel~c)], for the $\sigma$ polarized incident light,
the $0^{\circ}$ spectrum has smaller
L$_3$/L$_2$ ratio because of the contribution from the  $\sigma  \rightarrow \sigma'$ polarization which
is significant only at the L$_2$ edge.  In the case of the $\pi$ incident polarization, the branching ratio
slightly increases for the unpolarized situation, as the $\pi \rightarrow \pi'$ contribution at the L$_3$ edge
adds up to the main peak of the $\pi \rightarrow \sigma'$ polarization, while the L$_2$ contribution causes
appearance of a new small feature in the unpolarized spectrum.

Considering only the situation c) (no outcoming polarization analysis), the orbital spectra calculated
for the $\alpha=0^\circ$ and $\alpha=45^\circ$ structures show similar agreement with the
experimental spectrum in Fig.~\ref{fig:0oo}c). In this situation, one cannot discriminate between
the two cases. Very recently, however, experiments with a polarization analysis were performed
on \lsmo~ (Ref.~\onlinecite{StaScaMul05}), and no detectable orbital scattering was observed for the 
 $\sigma  \rightarrow \sigma'$ channel. Based on this observation, we conclude that the spin orientation
in \lsmo~ should correspond to an angle $\alpha$ equal to, or very close to, the value of 45$^\circ$, since
a measurable $\sigma  \rightarrow \sigma'$
component\cite{note3} should be present otherwise (see also Appendix A).

The magnetic spectra for the two CE structures  are compared in Fig. \ref{fig:AF}. 
The main difference between the   $\alpha = 0 ^\circ$  and $\alpha = 45^\circ$  cases 
is the decrease of the L$_3$/L$_2$ branching ratio in the case  $\alpha = 0 ^\circ$.
Otherwise, the spectra have a similar shape. We note that $\alpha = 45^\circ$ is also
the value which was used in Ref.~\onlinecite{WilStoBea05} to examine the effect of the Jahn-Teller distortion
on the orbital and magnetic spectra of \lsmo.
\begin{figure}[ht]
  \begin{center}
  \includegraphics[width=7.3cm, angle=270]{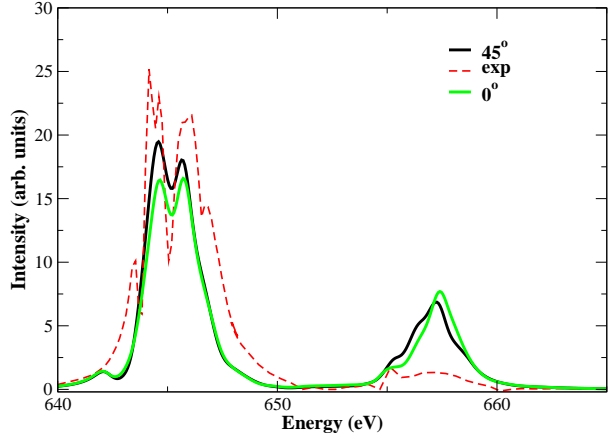}
  \caption{ (Color online) Theoretical magnetic order spectra of the CE structures for the 
two different angles enclosed by the spin and ${\vec b}$ axes: $\alpha = $45$^{\circ}$ (black line) and 
0$^{\circ}$ (light green line). 
The experimental spectrum (dashed red line) is
reproduced from Ref. \onlinecite{WilStoBea05}. }
  \bigskip
  \label{fig:AF}
  \end{center}
\end{figure}
The experimental orbital and magnetic scattering spectra and 
fits for \lsmo~ have been presented and discussed in 
Ref. \onlinecite{WilStoBea05}.

 Our finding that $\alpha\approx45^\circ$ differs from the
estimate of Staub {\it et al.}\cite{StaScaMul05}, who predicted
$\alpha = 10\pm5^\circ$, based on a fit focused on the  contribution of minority
ferromagnetic domains to the observed orbital scattering.
We note, however, 
that the angle $\alpha$ in the minority ferromagnetic domains and 
 in the majority CE domains may not be necessarily the same.

\begin{figure}[ht]
  \begin{center}
  \includegraphics[width=7.0cm, angle=270]{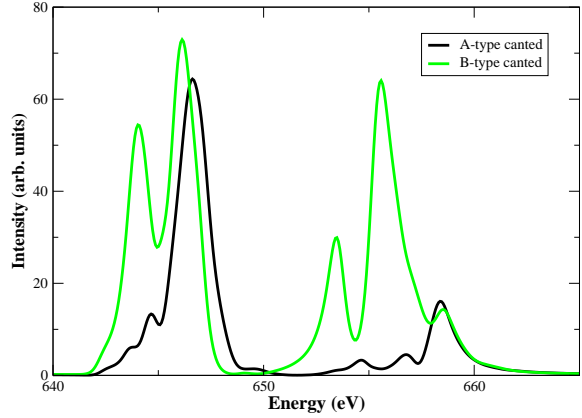}
  \caption{ (Color online) Theoretical spectra of the orbital scattering for 
the A-type (black line) and B-type (light green line) 
90$^\circ$-spin-canted structures, displayed in Fig.~\ref{fig:canted}.}
  \bigskip
  \label{fig:OO_canted}
  \end{center}
\end{figure}

To illustrate further the importance of taking into account the proper spin arrangement (and all
four different scattering sites), we
calculated the orbital scattering from the canted spin structures, in Fig.~\ref{fig:canted}. 
The orbital spectra for the
two canted structures are displayed in Fig.~\ref{fig:OO_canted}. It is clear that these two configurations
yield spectra largely different from the experimental and theoretical spectra for $\alpha = 45^\circ$, 
and that the calculation of the orbital scattering in \lsmo,
as shown in Fig.~\ref{fig:0oo},
should not be approximated by the consideration of only one site of high symmetry, A or B. 
At the same time, even from a quick inspection of  Fig.~\ref{fig:OO_canted},
one readily concludes that the two spectra are significantly different from each other. Yet, the only
difference in their calculation was the spin arrangement. 
From this, we conclude that changing the spin arrangement can have a drastic influence on the orbital
spectrum. This must be related, of course, in some way to 
the spin-orbit interaction. To  better identify the origin of this effect, 
we performed calculations
with either (i) the $d$ spin-orbit parameter set to zero or
(ii) the $p$-$d$ interaction set to zero. Calculations (i) 
yielded only minute changes to the spectra. Calculations (ii), instead, yielded major changes, 
resulting in two virtually identical spectra for the A and B canted structures. 
Therefore, we conclude that the sensitivity to the spin direction comes from the combined effect of the spin-orbit interaction 
in the $p$-shell and the $p$-$d$ Coulomb interaction, through which
$p$-electrons feel the crystal field \cite{note5}. 

\section{Effect of different types of spin misorientations on the orbital spectrum in the paramagnetic phase}

\begin{figure}[h]
  \begin{center}
  \includegraphics[width=7.3cm, angle=0]{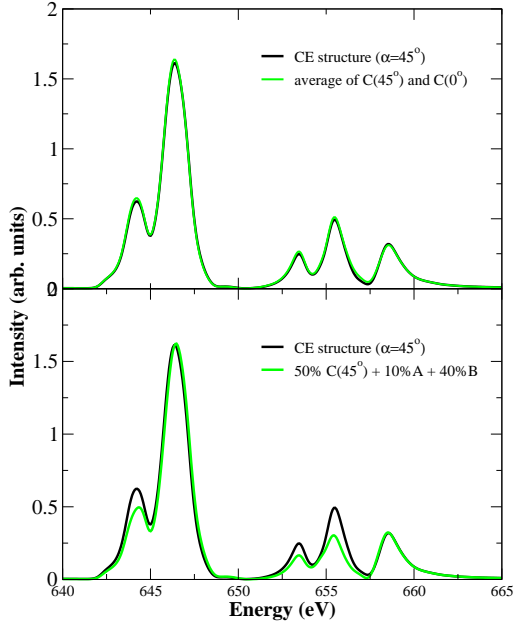}
  \caption{ (Color online) Comparison of the theoretical orbital scattering 
spectra for  $\alpha = 45^\circ$ (black line), 
and for configurations with different types of spin misorientations
in the paramagnetic case (light green line), described in the text. }
  \bigskip
  \label{fig:paramagnetic}
  \end{center}
\end{figure}
With the increase of temperature T above T$_N$, spin ordering
is first lost along the c-axis, while the in-plane correlations reduce
gradually until  the charge and orbital ordering temperature T$_{CO}$= 230~K, at which they vanish completely
\cite{SteHilWil96}. 
Based on the results from the previous sections, we discuss the influence of  
various spin misorientations on the orbital spectrum
in the paramagnetic phase for temperatures  T$_N \leq T \ll T_{CO}$, when there are still
correlations inside the $ab-$plane.  Randomness is expected to be mostly in the 
spin orientation between the planes, while the
antiferromagnetic ordering inside the planes is assumed to be less influenced.
 In the present work we do not consider any change in the orbital configurations within
the crystal.
We calculate the scattering contributions  at ${\vec q}=$\oo~
from CE $ab$-planes with different spin orientations for the 
angles belonging to the class of $\alpha=45^\circ + n\times 90^\circ$ ($n$ integer), which we 
defined as C(45$^\circ$), and
the class of  $\alpha=0^\circ+n\times 90^\circ$, C(0$^\circ)$, {\it i.e.} for
 $\alpha =$ 0$^\circ$, 45$^\circ$, 90$^\circ$, 
135$^\circ$, 180$^\circ$
 225$^\circ$, 270$^\circ$ and 315$^\circ$.
These planes are stacked in a random order, but assuming an equal proportion of each of them.
In this case, the scattering amplitude is given by the average of the scattering amplitudes of
the $\alpha=45^\circ$, 135$^\circ$, 0$^\circ$, and  $90^\circ$  cases. The corresponding 
orbital spectrum  is shown in panel~a) of Fig.~\ref{fig:paramagnetic} 
for the incident $\sigma$ polarization.
We see that the spectrum does not change with respect to the spectrum
obtained for the CE phase ($\alpha = 45^\circ$). This is because the $\sigma  \rightarrow \pi'$
scattering amplitudes for $\alpha = 0^\circ, 45^\circ, 90^\circ$, and 135$^\circ$ are all equal, and with
an equal amount of $\alpha=0^\circ$ and  $\alpha=90^\circ$ layers, the atomic scattering amplitude for the 
 $\sigma  \rightarrow \sigma'$ scattering component vanishes, reducing the spectrum
to the  $\sigma  \rightarrow \pi'$ scattering component.
Furthermore, assuming also a loss of coherence in the spin orientation between ferromagnetic
chains within the $ab$-plane and a random occurrence of the angles belonging to the classes
C(45$^\circ$) and C(0$^\circ$) for the ferromagnetic chains, the scattering amplitude can
also be shown to be that corresponding to the average of the $\alpha=0^\circ$, $\alpha=90^\circ$,
 and $\alpha = 45^\circ$ amplitudes. 

Hence, to produce a significant change in the orbital scattering, it is necessary to
consider other changes (with a higher degree of misorientation) in the spin configurations, such
as canted ``ferromagnetic'' chains.  We find, however, that an
 equal amount of A and B-type canted configurations 
yields a spectrum identical to the one for the CE phase, and that it is necessary to 
consider a different amount of
the two canted structures to change the spectrum. 
The orbital scattering spectrum for the case when 50 \% of the sites corresponds
to the configurations described with C(45$^\circ$), 10 \% to the A-type canted and 40 \% to the
B-type canted configurations is shown in b)~panel of Fig.~\ref{fig:paramagnetic}. The spectrum
is characterized by the reduction of intensity of the first two peaks at the L$_2$ edge, as well as of the
low-energy peak at the L$_3$ edge. 
The experimental
data of Wilkins {\it et al.}\cite{WilSpeHat03} indicate a large change of intensity at the
L$_3$ peak, a weaker one for the two low-energy peaks at the L$_2$ edge, and
virtually no change for the high-energy L$_2$ peak. A qualitatively similar trend can be deduced
from comparing the orbital spectra at 40~K and 160~K, measured by
Staub {\it et al.}\cite{StaScaMul05}, for which the largest intensity reduction
occurs at the L$_3$ edge and the low-energy peaks at the L$_2$ edge, while the intensity of
the highest-energy feature changes much less. 
A large change in the intensity of the main peak at the L$_3$
edge of the orbital spectrum at the N\' eel temperature could result from
the magnetic scattering of the minority ferromagnetic domains,
which would disappear above $\sim$ T$_N$\cite{StaScaMul05}. 
A similar effect, {\it i.e.} a significant reduction
of the main peak at the L$_3$ edge and almost negligible changes of the other peaks,
could also result from a decrease in 
the Jahn-Teller distortion\cite{WilStoBea05}. 
However, the change in intensity of the other peaks is difficult to explain with the
above effects, and Fig.~\ref{fig:paramagnetic}b indicates that they could be 
accounted for by a change in the spin configuration.

\section{Conclusions}
In this work, we analyzed the dependence of the scattering spectra on the spin configuration in \lsmo
 for a fixed orbitally ordered configuration.
We considered an arbitrary spin direction with respect to the local crystal field axis, yielding a low
local symmetry, which necessitated a general scattering formula written for the C$_i$ point group.
We generalized the Hannon-Trammel formula for the atomic scattering amplitude
by a systematic expansion in invariants of decreasing symmetry up to C$_i$ and implemented it 
in the framework of atomic multiplet calculations in a crystal field. 
We discussed in detail the cases of four different spin structures. 

We first considered two antiferromagnetic
CE structures with spins oriented either along the ${\vec b}$ axis or along an axis at an angle $\alpha=45^\circ$
relative to the ${\vec b}$ axis, within the $ab$-plane.  The latter structure  has the
highest symmetry (all sites are related by symmetry operations) and has only the
$\sigma  \rightarrow \pi'$ orbital scattering for $\sigma$ incident light
 which is confirmed by recent experiments which
found no $\sigma  \rightarrow \sigma'$ scattering component. For the consideration of the 
CE structure with spins aligned with the lattice vector ${\vec b}$  
($\alpha=$ 0$^\circ$), one needs to include an equal amount of sites which have spins either
at 0$^\circ$ or at 90$^\circ$
with respect to the local crystal field axis in order to keep 
a collinear spin configuration. 
Orbital scattering in the $\sigma  \rightarrow \pi'$ channel 
for the cases $\alpha=$ 45$^\circ$ and $\alpha=$ 0$^\circ$ yields identical spectra, but in the latter case 
the scattering also takes place in the $\sigma  \rightarrow \pi'$ channel, in contrast to the experimental
situation. 
Similarly, for the incident $\pi$ polarization, there is a nonvanishing  $\pi \rightarrow \pi'$
contribution, and the spectrum of the $\pi \rightarrow \sigma'$ channel is identical to the
one for  $\alpha=45^\circ$. For the magnetic scattering, the spectrum of the configuration with
$\alpha=45^\circ$ displays a somewhat larger L$_2$/L$_3$ ratio. From the comparison with experiment, we conclude 
that in the \lsmo~ CE phase the angle between the spins and ${\vec b}$ axis is equal, or very close to,
45$^\circ$.

We then also examined two 90$^\circ$-spin-canted structures obtained by considering only sites having the
spin direction either at 0$^\circ$ or at 90$^\circ$ from the local crystal field axis. The spectra differ
greatly in the two cases, and also differ from those obtained for the antiferromagnetic CE structures.
 From the consideration of these structures, we thus conclude that the spin configuration
can have a drastic influence on the orbital spectrum. The largest differences caused by the spin
configuration on the orbital spectra are observed in the presence of maximal disproportionation between
spins parallel and perpendicular to the local crystal field axis.

Finally, we also discussed the changes observed with temperature, in the orbital spectrum of the
paramagnetic phase, considering the effect of various types of spin misorientations with respect to the
CE structure. We first examined the effect of an incoherence in the 
spin orientation between the $ab$-planes and
then also between the ferromagnetic chains within the $ab$-planes, considering ferromagnetic chains with spin
directions corresponding to $\alpha=0^\circ$, 45$^\circ$, 90$^\circ$, 135$^\circ$, 180$^\circ$, 225$^\circ$,
270$^\circ$, and 315$^\circ$ in equal proportion. 
This was found to have a negligible effect on the orbital spectrum. We then
examined the effect of a disorder in the spin direction within the ferromagnetic chains, considering
the contributions of different types of 90$^\circ$-spin-canted chains with more spins parallel than
perpendicular to the local crystal field axis. This resulted in a spectrum
with no change in the highest-energy peak, both at the L$_2$ and L$_3$ edges, but a
reduced intensity of all other peaks. In view of explaining the measured temperature
dependence of the orbital scattering intensity,  the changes of the spectrum caused by the
spin misorientations can complement the effects 
which would arise due to the presence of possible minority ferromagnetic domains or a reduction 
in the Jahn-Teller distortion, which would mostly affect the highest-energy peak at the
L$_3$ edge.

\appendix 

\section{ Orbital and magnetic scattering for an arbitrary angle between the spin and lattice axes}

Previously, we discussed the scattering in the CE structure when the in-plane spin and lattice axes 
enclose angles of 0$^\circ$ or 45$^\circ$. At this point, we present the expression for the orbital 
scattering amplitude in the case of an arbitrary angle $\alpha$ between the spin axis and lattice vector ${\vec b}$ 
for  $\sigma$-polarized incident light:
\begin{equation}
\begin{array}{lll}
f^{{\rm OO, \sigma inc}}_{\rm res} &=& \left\{ \begin{array}{ll} \sin^2\beta \cos^2\phi& \\  \frac{1}{2}(\sin^2\beta \sin\theta \sin2\phi   -\sin 2\beta \cos\theta \cos\phi) & \end{array}\right\}   2(F_{0;0}^{A+}- F_{0;0}^{B+}) + \\ 
& & \left\{ \begin{array}{ll}\cos^2\beta \cos^2\phi + \sin^2 \phi & \\ \frac{1}{2}[\sin2\phi \sin\theta(\cos^2\beta-1)+\sin 2 \beta \cos\theta \cos\phi]  & \end{array}\right\} (F_{-1;-1}^{A+}+F_{1;1}^{A+}-F_{-1;-1}^{B+}-F_{1;1}^{B+}) + \\
& & \left\{ \begin{array}{ll} \cos^2\beta \cos^2\phi - \sin^2 \phi  & \\\frac{1}{2}[\sin2\phi \sin\theta(\cos^2\beta +1)+\sin 2 \beta \cos\theta \cos \phi]   & \end{array}\right\}  (-F_{-1;1}^{A+}-F_{1;-1}^{A+}+F_{-1;1}^{B+}+F_{1;-1}^{B+})  
\\ & & - \frac{1}{\sqrt{2}} \left\{ \begin{array}{ll}\sin2\beta \cos^2\phi  & \\ \frac{1}{2}\sin2\beta \sin\theta \sin 2\phi - \cos2\beta \cos \theta \cos\phi & \end{array}\right\}  (F_{1;0}^{A+}-F_{0;-1}^{A+}-F_{-1;0}^{A+} + F_{0;1}^{A+} \\
& & -F_{1;0}^{B+}+F_{0;-1}^{B+}+F_{-1;0}^{B+} - F_{0;1}^{B+}),
\label{eq:arb_alpha}
\end{array}
\end{equation}
where we used $\beta=45^\circ - \alpha$ and the definitions of $\theta$ and $\phi$ from Fig~\ref{fig:scatter}.
We remind the reader that the $F_{m,m'}$ spherical tensor components 
are defined with respect to the spin axis, and that their values
change with the angle $\alpha$. We also note that there is an additional term in the scattering amplitude
(the last term in the above equation), with respect to  Eq.~\ref{eq:OOscat0-90}, which does 
not vanish in the general case. In Cartesian coordinates this term is proportional to $F_{xz}+F_{zx}$.
Considering, {\it e.g.}, the azimuthal angle $\phi=0^\circ$, we note that 
there is a $\sigma \rightarrow \sigma'$ scattering component for all angles  
$0\leq \alpha \leq 90^\circ$, except for $\alpha=45^\circ$.

In the case of magnetic scattering, there are no additional terms
(compared to Eqs.~\ref{eq:AF_45} and \ref{eq:AF_0}) contributing to the scattering. The scattering always
takes place only in the $\sigma \rightarrow \pi'$ channel.
 
\section{ Linear expansion of F$_{m,m'}$ around $\alpha_{A(B)}=45^\circ$}

If we denote the angle between the spin and  the local crystal field axis at site A (B)  as 
$\alpha_A$ ($\alpha_B$), then the two angles in Fig.~\ref{fig:structure}, Case II,
 are related by: $\alpha_A=90^\circ - \alpha_B$.
We will expand the tensor $F_{m,m'}^A$ defined in Eq.~\ref{eq:defineF}, as a function of 
the angle $\alpha_A$ around 45$^\circ$ in the following way:
\begin{equation}
F_{m,m'}^A(\alpha_A)= F_{m,m'}^A(45^\circ)+\frac{\delta F_{m,m'}^A(\alpha_A)}{\delta \alpha_A}\Big|_{\alpha_A=45^\circ}(\alpha_A-45^\circ) + O(\Delta \alpha_A^2).
\end{equation}
The same expansion is valid for  $F_{m,m'}^B$. Therefore, for the $F_{xx}^A$ and $F_{zz}^B$ 
from  Eq.~\ref{eq:OOscat0-90_cart}, we can write: 
\begin{equation}
F_{xx}^A(\alpha_A)= F_{xx}^A(45^\circ)+\frac{\delta F_{xx}^A(\alpha_A)}{\delta \alpha_A}\Big|_{\alpha_A=45^\circ}(\alpha_A-45^\circ)+ O(\Delta \alpha_A^2),
\label{eq:FxxA}
\end{equation}
where $x$ is the coordinate along ${\vec a}$ and $z$ the coordinate along ${\vec b}$.
We can also write:
\begin{equation}
\begin{array} {lll}
F_{zz}^B(\alpha_B)& &= F_{zz}^B(45^\circ)+\frac{\delta F_{zz}^B(\alpha_B)}{\delta \alpha_B}\Big|_{\alpha_B=45^\circ}(\alpha_B-45^\circ)+ O(\Delta \alpha_B^2) \\
& & = F_{xx}^A(45^\circ)+\frac{\delta F_{xx}^A(\alpha_A)}{\delta \alpha_A}\Big|_{\alpha_A=45^\circ}(45^\circ - \alpha_A)+ O(\Delta \alpha_A^2),
\label{eq:FzzB}
\end{array}
\end{equation}
using the fact that $\Delta\alpha_A=-\Delta\alpha_B$ and that $F_{xx}^A(45^\circ)=F_{zz}^B(45^\circ)$ and $\frac{\delta F_{xx}^A(\alpha_A)}{\delta \alpha_A}\Big|_{\alpha_A=45^\circ}=\frac{\delta F_{zz}^B(\alpha_B)}{\delta \alpha_B}\Big|_{\alpha_B=45^\circ}$,
because the atomic scattering tensor $F^A(\alpha_A=45^\circ+\Delta \alpha_A)$ at the A site is equivalent to the atomic
scattering tensor $F^B(\alpha_B=45^\circ+\Delta \alpha_A)$ at the B site through a rotation of 180$^\circ$ about
the ${\vec a}+{\vec b}$ axis.  
In the intensity expression, Eq.~\ref{eq:OOscat0-90_cart}, $F_{xx}^A$ and
$F_{zz}^B$ are added up. From Eqs.~\ref{eq:FxxA} and \ref{eq:FzzB}, we see that the
two first order terms  will have the same magnitude and opposite signs. The same can be shown for the
remaining terms from Eq.~\ref{eq:OOscat0-90_cart},  $F_{zz}^A$ and $F_{xx}^B$, thus
proving that the first order terms in the expression for the intensity all vanish, leaving only the
zeroth order term equal to the  $45^\circ$
contribution. We also tested the assumption of a linear dependence of $F_{m,m'}$ on the angle around 45$^\circ$ by
checking numerically the linear-order result:
\begin{equation}
F_{xx}^A(\alpha_A=0^\circ)+F_{xx}^A(\alpha_A=90^\circ)=2 F_{xx}^A(\alpha=45^\circ),
\end{equation}
and confirmed that the dependence $F_{m,m'}(\alpha)$ is described well (within the numerical noise) 
with the expansion up to 
the first order.

\acknowledgments
We are grateful to P. Carra for helpful discussions and his assistance 
in learning how to use the Cowan and ``Racah'' codes. We acknowledge stimulating 
discussions with G. Trimarchi.

\bibliography{spin}

\begin{thebibliography}{34}
\bibitem[{not({\natexlab{a}})}]{note2}
\bibinfo{note}{There is experimental indication for the spin direction in the
  [210]-direction \cite{SteHilWil96}. However, in the present study, we will
  consider two different spin orientations and, on the basis of theoretical
  considerations and indirect experimental evidence, propose the alternative
  spin arrangement.}
\bibitem[{not({\natexlab{c}})}]{note1}
\bibinfo{note}{The matrix elements from Eq. \ref{eq:defineF} are evaluated for
  the spherical symmetry case, yielding either a finite real or zero value. By
  repeated application of the Wigner-Eckart theorem (multiplication with the
  corresponding 2-j and 3-j coefficient) these matrix elements are ``lowered''
  to the final symmetry. All 2-j and 3-j coefficients in the applied symmetry
  branching are real and, therefore, the final matrix elements and F's itself
  are real.}
\bibitem[{not({\natexlab{d}})}]{note3}
\bibinfo{note}{The atomic scattering amplitudes of the configurations with
  $\alpha=0^\circ$ and $\alpha=90^\circ$ are the same for the $\sigma
  \rightarrow \pi'$, and of opposite sign for the $\sigma \rightarrow \sigma'$
  scattering channel. The coexistence of both configurations in domain
  structures might lead to the conclusion that also in this case the $\sigma
  \rightarrow \sigma'$ scattering component vanishes. However, we note that
  considering also the domains corresponding to the two equivalent ${\rm
  A-B-A-B-\ldots}$ and ${\rm B-A-B-A-\ldots}$ arrangements of orbitally ordered
  planes along ${\vec a}$$+{\vec b}$, which have, in principle, identical
  probability to occur, the $\alpha=0^\circ$ and $\alpha=90^\circ$ domains are
  expected to give contributions to the scattering amplitude which are out of
  phase with each other. Their signals, therefore, should be added on the level
  of intensity, thus keeping the same proportion of the $\sigma \rightarrow
  \sigma'$ scattering component in the total signal, as in the case of
  $\alpha=0^\circ$ (or $\alpha=90^\circ$).}
\bibitem[{not({\natexlab{e}})}]{note5}
\bibinfo{note}{When the $p$-$d$ interaction is vanishing, and the $d$
  spin-orbit interaction is zero, the spin of the $d$ electrons, ${\vec S_d}$,
  and its rotations commute with the Hamiltonian. As a result, the scattering
  amplitude does not depend on the spin direction. On the other hand, when the
  $p$-$d$ interaction is not vanishing, ${\vec S_d}$ and its rotations do not
  commute anymore with the Hamiltonian. Furthermore, we note that, for the
  ground state, the total spin of the $p$- and $d$- electrons, ${\vec
  S_{tot}}$, also coincides with the $d$- electron spin. However, the
  Hamiltonian is not invariant under the rotations of the total spin (because
  of the $d$- crystal field and the $p$ spin-orbit interaction). Therefore, in
  the presence of $p$-$d$ interaction, the scattering amplitude depends, in
  general, on the direction of the spin of the ground state.}
\end{thebibliography}

\end{document}